\documentclass[twocolumns]{aa}

\usepackage{txfonts}
\usepackage{graphicx}	
\usepackage{amsmath}	

\usepackage{amssymb}	
\usepackage{xcolor}     
\usepackage{soul}       
\usepackage{array}
\usepackage{comment}
\usepackage{csquotes}
\usepackage{subfig}
\usepackage{float}
\usepackage{multirow}
\usepackage{makecell}
\usepackage{wrapfig}
\usepackage[export]{adjustbox}
\usepackage[colorlinks=true,linkcolor=blue,citecolor=blue, urlcolor=blue]{hyperref}
\usepackage{natbib}
\bibpunct{(}{)}{;}{a}{}{,} 

\usepackage{fontawesome}  

\definecolor{fredcolor}{RGB}{204, 204, 0}
\definecolor{changesgreyblue}{RGB}{78, 129, 166}
\definecolor{austincolor}{HTML}{00D4B1}

\newcommand{\lcdm}{$\Lambda$CDM\xspace}
\newcommand{\herculens}{\textsc{Herculens}\xspace}

\newcommand{\molet}{\textsc{MOLET}\xspace}

\newcommand{\op}[1]{\ensuremath{\boldsymbol{\mathsf{#1}}}\xspace}
\newcommand{\mat}[1]{\ensuremath{{\rm \bf #1}}\xspace}

\newcommand{\lensingop}{\op{L}}
\newcommand{\convop}{\op{B}}

\begin{document}


\title{Modeling lens potentials with continuous neural fields\\in galaxy-scale strong lenses}

\titlerunning{Lens potential modeling with neural networks}

\author{L.~Biggio\inst{\ref{eth}}\fnmsep\thanks{luca.biggio@inf.ethz.ch}, G.~Vernardos\inst{\ref{epfl}}, A.~Galan\inst{\ref{epfl}}, A.~Peel\inst{\ref{epfl}}}

\institute{
Data Analytics Lab, Institute of Machine Learning, Department of Computer Science, ETHZ, Switzerland \label{eth},
\goodbreak
\and
Institute of Physics, Laboratory of Astrophysics, Ecole Polytechnique 
F\'ed\'erale de Lausanne (EPFL), Observatoire de Sauverny, 1290 Versoix, 
Switzerland \label{epfl}
}

\abstract{
Strong gravitational lensing is a unique observational tool for studying the dark and luminous mass distribution both within and between galaxies. Given the presence of substructures, current strong lensing observations demand more complex mass models than smooth analytical profiles, such as power-law ellipsoids. In this work, we introduce a continuous neural field to predict the lensing potential at any position throughout the image plane, allowing for a nearly model-independent description of the lensing mass. We apply our method on simulated Hubble Space Telescope imaging data containing different types of perturbations to a smooth mass distribution: a localized dark subhalo, a population of subhalos, and an external shear perturbation. Assuming knowledge of the source surface brightness, we use the continuous neural field to model either the perturbations alone or the full lensing potential. In both cases, the resulting model is able to fit the imaging data, and we are able to accurately recover the properties of both the smooth potential and of the perturbations. Unlike many other deep learning methods, ours explicitly retains lensing physics (i.e., the lens equation) and introduces high flexibility in the model only where required, namely, in the lens potential. Moreover, the neural network does not require pre-training on large sets of labelled data and predicts the potential from the single observed lensing image. Our model is implemented in the fully differentiable lens modeling code \herculens.
}


\maketitle

\section{Introduction}

On large scales of the Universe, typically beyond 10 Mpc, the Lambda Cold Dark Matter (\lcdm) model, and in particular its Dark Matter (DM) component, successfully explains the formation of galaxies through the hierarchical gravitational collapse of matter \citep[e.g.][]{Toomre1972,Dubinski1994,Springel2006}. Below 10 Mpc, the gravitational interactions between cold DM and baryons perpetually reshape the mass distribution in between and within galaxies. Many complex mechanisms are thought to affect the mass content of galaxies, including gas outflows from active galactic nuclei \citep[e.g.][]{ZubovasKing2012}, supernova feedback \citep[e.g.][]{Scannapieco2008}, and tidal stripping from ongoing mergers \citep[e.g.][]{BarnesHernquist1996}. As certain challenges remain unresolved within the cold DM paradigm [e.g. the ``cusp-core problem'' \citep{Moore1994,deBlock2010}, the ``missing satellites problem'' \citep{Moore1999,Klypin1999}, and the ``too-big-to-fail problem'' \citep{BoylanKolchin2011,Papastergis2015}], it is crucial to improve techniques that characterize the dark matter distribution in galaxies, and to compare it to theoretical predictions.

Strong gravitational lensing offers a direct probe of the total mass distribution within galaxies. In the galaxy-galaxy regime, this phenomenon arises when two galaxies located at different redshifts are aligned along our line of sight: the mass of the foreground lens galaxy causes a deflection of the light emitted by the background source galaxy. From the observer's point of view, multiple images of the source galaxy are visible and potentially form lensed arcs. Although magnified and highly distorted, these arcs encode significant information about the underlying mass distribution of the lens galaxy. The mass distribution is usually described through a lens potential for which simple functions, such as elliptical power-law profiles, often lead to a sufficiently good fit to the observation \citep[e.g.][]{Shajib2021}. However, the true lens potential of galaxies is known to be more complex and can feature ellipticity twists or gradients, a bar component, frozen shocks, or include faint satellite galaxies. Moreover, populations of DM subhalos along the line of sight also contribute to the observed lensing distortions. Therefore, models that go beyond a simple smooth component are required to capture the complexity of the lens galaxy mass distribution, including its dark component.

Strong lens modeling requires solving an under-constrained problem, notably because the unlensed source light distribution is unknown and its multiple images only span a limited fraction of the lens extent. To address this difficulty, smooth analytical profiles have been widely used to regularize the model of the lens potential. If one observes clear signs of structures perturbing the smooth lens potential, such as luminous satellite galaxies or shear effects from a nearby galaxy cluster, it is possible to inject prior knowledge in the model by using well-motivated analytical profiles. However, in most cases, the signatures of complex mass distributions are less obvious, and can range from dark subhalos to high-order moments in the lens potential. Direct reconstruction methods are preferable in such cases, as they relax the strong assumptions on the underlying mass distribution that fully analytic strategies impose. An example is the gravitational imaging technique developed by \citet{Koopmans2005}, which reconstructs both perturbations of the smooth lens potential and of the source light on pixelated grids, based on a first-order expansion of the lens equation. However, the success of the technique can strongly depend on the choice of regularization used to reconstruct the pixelated perturbations \citep{Vernardos2022}. Recently, \citet{Galan2022} introduced a novel multi-scale regularization technique based on wavelet transforms and sparsity priors to reconstruct different types perturbations on a pixelated grid.

Aside from analytical and pixelated models, there has been growing interest in using deep learning to model the lens potential. Such approaches can offer much lower computation times compared to traditional methods, as well as the ability to learn complex features in the data. Most studies so far have focused on training deep convolutional neural networks (CNNs) to predict parameter values of usual analytical lens models \citep[e.g.][]{Hezaveh2017,Pearson2019}, including the estimation of parameter uncertainties with approximate Bayesian neural networks \citep[BNNs, e.g.][]{PerreaultLevasseur2017,WagnerCarena2021}. The generation of realistic training sets is very challenging, but progress has been made recently with the use of real images of source and lens galaxies \citep{Schuldt2021}. A common feature of these approaches is that they do not explicitly rely on the well-understood physics of gravitational lensing, namely the lens equation. Instead, they implicitly learn a representation of lensed features through hundreds of thousands of example realizations from a training set.

Deep neural networks have also been used for the detection of DM subhalos \citep{Alexander2020b}, the recovery of their angular positions \citep{Coogan2020,YaoYuLin2020,Ostdiek2022}, and distinguishing among different types of DM \citep[e.g.][]{Alexander2020a}. Since considering the effect of populations of subhalos allows probing lower subhalo masses \citep{Hezaveh2016a}, similar techniques have been used to frame the weighing of subhalos as a classification problem \citep{Varma2020}, to constrain the power-spectrum of lens potential perturbations with statistical uncertainties \citep{Vernardos2020}, and to constrain the DM subhalo mass function \citep{Brehmer2019,DiazRivero2020,WagnerCarena2022}. While these deep learning approaches probe many facets of the lens complexity problem, they remain limited by the need for large and realistic training sets (in particular for CNNs), which can be challenging to design and generate. Furthermore, as deep learning-based techniques have not yet been applied to real strong lensing data, it is still unclear whether assumptions in the training sets might limit their application to more complex systems.

Recently, a novel class of approaches have been explored by several authors that combine deep learning with physics-based modelling. The goal of such \emph{hybrid} approaches is to retain known physics explicitly while still achieving high model complexity and low computation time. This new paradigm has resulted in a number of successful methods across a wide range of fields, such as robotics \citep{robotics}, quantum mechanics \citep{Hermann_2020}, molecular dynamics \citep{doerr2021torchmd}, and fluid-dynamics \citep{thuerey2021physics}. Similar in spirit is the concept of physics-informed neural networks \citep{raissi2019physics,pinns}, where machine learning is used to fit the data while satisfying constraints imposed by complex partial differential equations from physics. 

In the context of strong lensing, \citet{Chianese2020} introduced a modeling method where the source light is generated on a pixelated grid by a pre-trained variational auto-encoder (VAE) and then combined through the lens equation with an analytic description of the lens potential to produce an image of the lensed source. This is made possible using differentiable programming, where all model components, namely the VAE (by construction), the analytic profiles, and the lens equation are written as automatically differentiable functions, allowing for the efficient optimization of parameters via gradient descent. In a similar spirit, \citet{Adam2022} trained neural networks on real galaxy images and cosmological simulations and use them as custom priors to reconstruct the source light and the lens potential, respectively. Differentiable programming has also been used in \citet{Karchev2022} to model lensed sources with Gaussian processes and variational inference, and \citet{Gu2022} constructed a GPU-accelerated Bayesian framework for fast modeling and parameter inference applied to large samples of lenses.

Another recent effort toward fully differentiable strong lens modeling is \herculens\footnote{\url{https://github.com/austinpeel/herculens}} \citep{Galan2022}, which combines analytical and pixelated models together with deep learning optimization methods into a single modular framework. \herculens allows for fast gradient-informed optimization over thousands of parameters, in particular the reconstruction of pixelated source light distribution and lens potential perturbations regularized with wavelets. It is based on JAX \citep{jax2018github} for performing high-performance automatic differentiation tasks.

\begin{figure}
    \centering
    \includegraphics[width=0.48\textwidth]{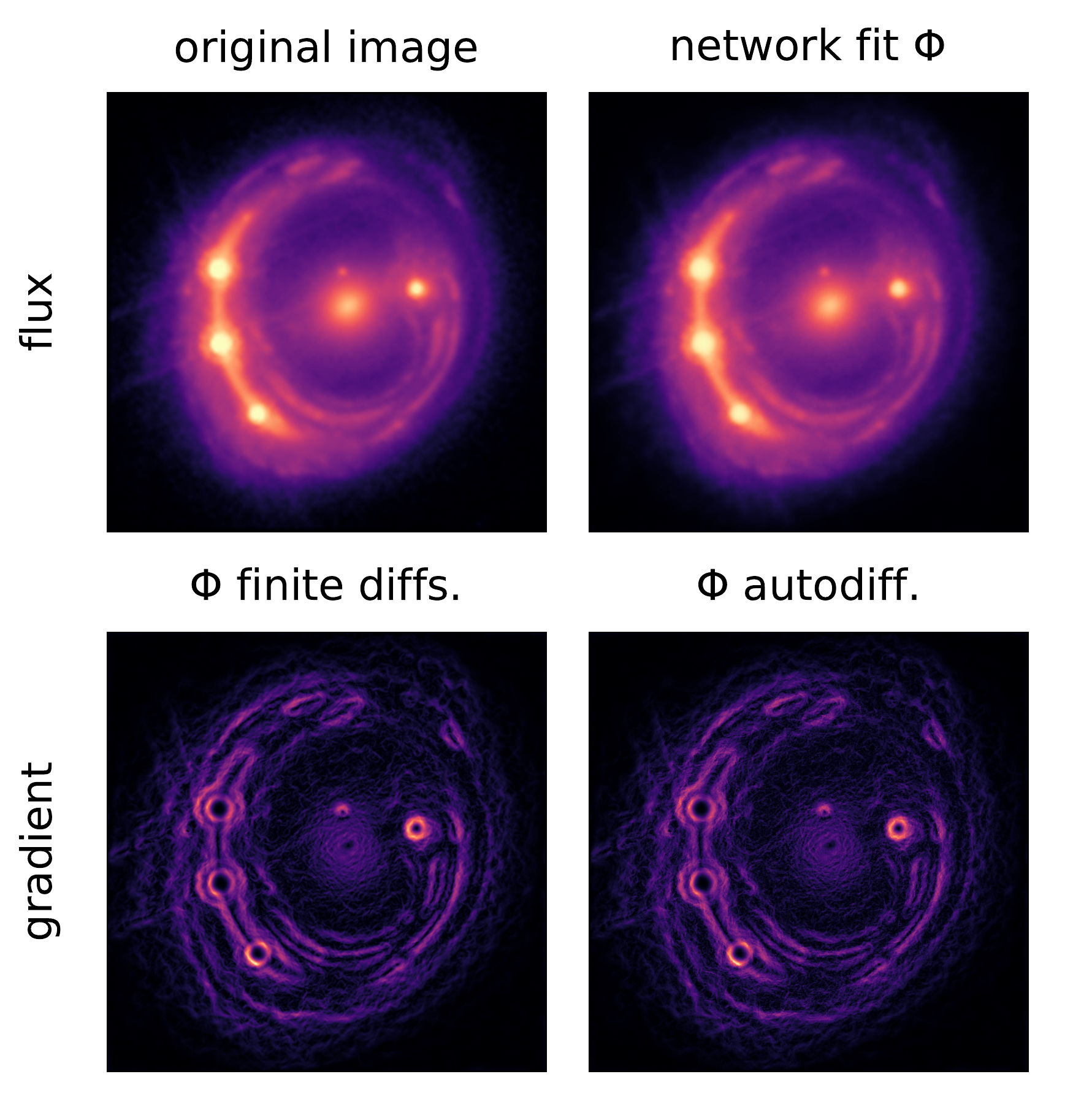}
    \caption{Neural network representation of the well-known strongly lensed quasar system RX\,J1131$-$1231, captured by the Hubble Space Telescope \citep[see e.g.][]{Suyu2013}. The network was trained to map pixel positions into pixel values and is able to capture fine details of both the original image and of its gradients down to a few percent. The magnitude of the gradient of $\Phi$, computed by finite differences and by automatic differentiation through the network, is shown for comparison.} 
    \label{fig:exampleSL}
\end{figure}

Following up the proof-of-concept work of \citet{Biggio2021neurips}, we introduce a new fully differentiable method to model the lens potential using a deep neural network that leverages the auto-differentiable capabilities of \herculens. In particular, we draw inspiration from the growing literature on implicit representation learning \citep{nerf,sitzmann2019siren,tancik2020fourier}, whose goal is to parameterize a generic signal in a continuous and differentiable fashion by means of a neural network. As an example of this concept, Fig.~\ref{fig:exampleSL} showcases the capacity of a simple fully-connected network to accurately reproduce the details in an image of a strong lens. We trained the network to map pixel positions into pixel values by minimizing the squared error between the network output and the data image. The network we present to model the lens potential operates similarly, but with the difference that the true target potential is unknown. It acts as a modular building block within the \herculens framework and interacts with the other simulation components according to the physics of strong lensing. In contrast to pixelated methods, our network is able to model the lens potential continuously at any position in the observed field of view and can equally be used to represent perturbations of a smooth profile or the full potential itself. We note that recently, a similar technique has been used to reconstruct the lensed source light distribution \citep{MishraSharma2022}.

We test our method in various realistic contexts, ranging from a fully smooth lens potential to perturbations on top of a smooth profile caused by a localized DM subhalo, populations of DM subhalos, or higher-order moments within the lens mass distribution. Despite the large number of parameters, automatic differentiation enables efficient (and exact) loss function gradient and higher-order derivative computations, facilitating rapid training and convergence to the best-fit solution. We emphasize that training in our case is performed with only the target lensing image, avoiding the typical requirement of large sets of labeled examples. Furthermore, our neural network implementation of the potential allows \textit{spatial} derivatives to be computed automatically, meaning deflection angles and other lensing quantities (e.g. convergence and shear) can be obtained with high accuracy.

The paper is organized as follows. We present our methodology in Sect.~\ref{sec:method}, which includes the elements of the strong lensing formalism required for our work. We also describe the framework of implicit representation learning as applied to modeling the lensing potential, along with details of the network architecture and training. In Sect.~\ref{sec:experiments}, we describe the experiments carried out on mock data to test our method. We present our experimental analysis in Sect.~\ref{sec:results} and assess the performance of the proposed approach. We conclude and discuss future extensions and research directions in Sect.~\ref{sec:conclusion}.

\section{Methodology}\label{sec:method}

\subsection{Strong gravitational lensing}\label{sec:strong_lensing}

Apparent distortions of a distant luminous source caused by the gravitational lensing of a foreground mass distribution are captured by the lens equation:
\begin{align}
    \label{eq:lens_equation}
    \vec\beta=\vec\theta-\vec\alpha(\vec\theta).
\end{align}
It is a mapping from lens plane coordinates $\vec\theta$ to (unlensed) source plane coordinates $\vec\beta$, both two-dimensional angular position vectors on the sky. The reduced deflection angle $\vec\alpha$ is in general a non-linear function of the lens plane position and can be derived from the projected lens mass density, namely
\begin{align}
    \label{eq:deflection_angle}
    \vec\alpha(\theta)=\nabla\psi(\vec\theta),
\end{align}
where $\psi$ is the lens potential in the limit of the thin lens approximation.

As gravitational lensing preserves surface brightness, a source plane image described by $\mathcal{I}(\vec\beta)$, for example the unlensed image of a galaxy, is mapped to the lens plane through
\begin{align}
    \label{eq:surface_brightness}
    \mathcal{I}(\vec\theta)=\mathcal{I}\Big(\vec\beta(\vec\theta)\Big).
\end{align}
This relation is continuous and holds for any light ray emanating from the source and arriving at the observer. In practice, however, our telescopes capture only discretized images of the sky and are subject to instrumental noise. We can thus postulate the problem in a discretized way so that the data (i.e. image) we observe, $\vec d$, is given by
\begin{align}
    \label{eq:linear_lens}
    \vec{d} = \convop\lensingop_{\psi}\vec{s} + \convop\vec\ell + \vec{n},
\end{align}
where $\vec{s}$ is the source light profile, $\lensingop_{\psi}$ is a discretized version of Eq.~(\ref{eq:lens_equation}) encoding the lens mapping, and $\vec\ell$ is the light profile of the lens. We use the subscript $\psi$ on $\lensingop_{\psi}$ to emphasize that the precise form the lensing operator takes depends (non-linearly) on the lens potential. $\convop$ is a blurring operator modeling the point spread function (PSF) of the instrument and representing the seeing conditions. The final term, $\vec{n}$, is additive noise, typically a combination of instrumental read-out noise and signal-dependent shot noise.

Our goal in this work is to recover the lens potential $\psi$ and source light $\vec{s}$ in Eq.~(\ref{eq:linear_lens}) given an observed image of strong lensing $\vec{d}$. We assume throughout that the PSF is known and constant across the field of observation. As it is not our focus here, we also assume that the lens light $\vec\ell$ has been accurately modeled and subtracted from the data in each case, although our modeling framework enables it to be straightforwardly included. Blending of the lens light with lensed source features can make disentangling the two difficult as well as introduce degeneracies. We therefore leave a full treatment of Eq.~(\ref{eq:linear_lens}) including lens light for future work.

Smooth lens mass distributions are often described by analytic profiles in terms of (dimensionless) convergence $\kappa$, where
\begin{align}
    \label{eq:convergence}
    \kappa(\vec\theta)=\frac{1}{2}\nabla^2\psi(\vec\theta).
\end{align}
For example, the singular isothermal ellipsoid (SIE) is commonly used to model the large-scale features of the lensing mass and is defined as a function $\kappa(\vec\theta)$. The profile has analytic formulas for the corresponding potential $\psi$ and deflection angle $\vec\alpha$ \citep[see, for example, Appendix B of][]{Galan2022}. Small-amplitude deviations from smoothness arising from various physical mechanisms (see Sect.~\ref{sec:experiments}) can be included as an additional analytic or pixelated component in the potential. In such cases, we can distinguish the smooth analytic part $\psi_\mathrm{sm}$ from the perturbations $\psi_\mathrm{pert}$ and write 
\begin{align}
    \psi = \psi_\mathrm{sm} + \psi_\mathrm{pert}.
\end{align}
The neural network model for $\psi$ that we explore in this work is general and can used to represent both perturbations on top of a smooth potential, namely $\psi_\mathrm{pert}$ alone, as well as the full potential $\psi$ itself.

\subsection{Neural networks for implicit representation}

\begin{figure*}
    \centering
    \includegraphics[width=\textwidth]{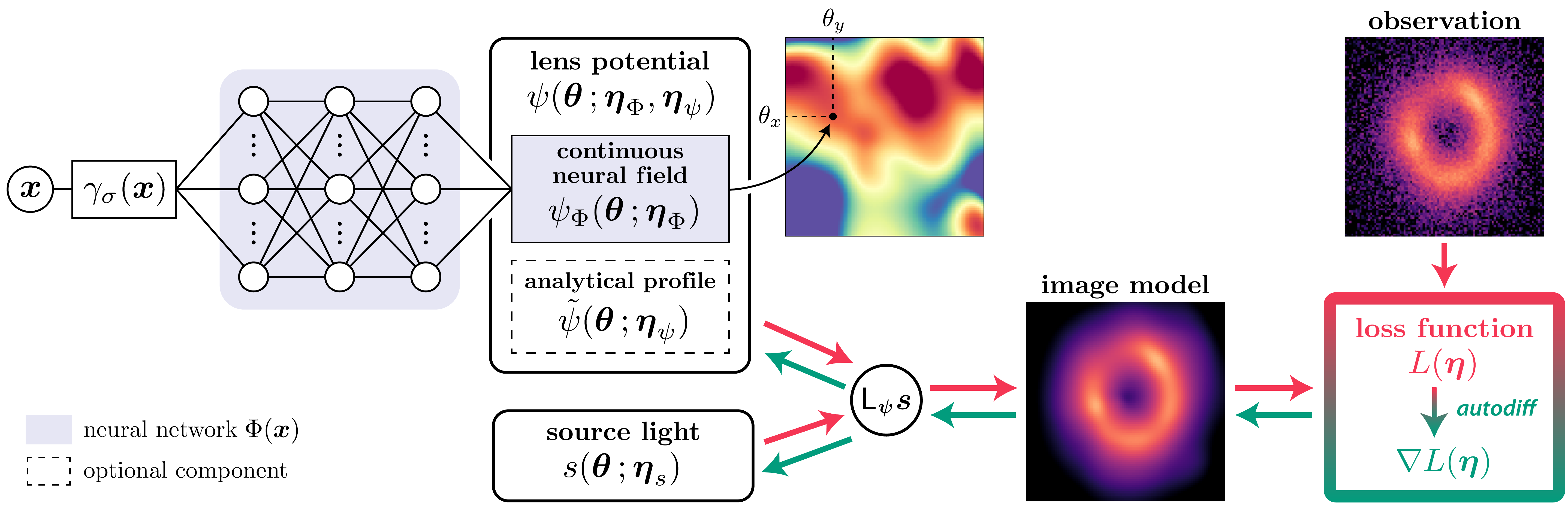}
    \caption{Flowchart of the proposed method. The lens potential is modeled with a continuous neural field that takes as input any position in image plane and outputs the value of the lens potential at that position. Alternatively, an analytical profile (e.g., a SIE) models the smooth component of the lens potential, while the neural network captures deviations from that smooth component. The input coordinates are first passed through a Fourier feature mapping ($\gamma_\sigma$) to increase the dynamic range of the recovered features. The different model components follow Eq.~(\ref{eq:linear_lens}) except for the blurring operator, omitted to avoid clutter. Since the model is fully differentiable, automatic differentation is used to compute the exact gradient of the highly non-linear loss function.}
    \label{fig:model}
\end{figure*}

Given an input vector $\vec{x}\in\mathbb{R}^d$, the action of the simplest deep neural network model, the multilayer perceptron (MLP), is the sequential application of affine transformations $T(\vec{x})$, defined by weight matrices $\mat{W}$ and biases $\vec{b}$, and an element-wise non-linear activation function $\phi$: 
\begin{align}
    \label{eq:mlp}
    \Phi(\vec{x})=\left(T^{(n_l)} \circ \phi \circ T^{(n_l-1)} \circ \ldots \circ \phi \circ T^{(1)}\right)(\vec{x}),
\end{align}
where
\begin{align}
    T^{(i)}(\vec{x})=\mat{W}^{(i)}\vec{x} + \vec{b}^{(i)}
\end{align}
is the $i$th transformation and $n_l$ is the number of layers. In the case of representing image values based on input coordinates, $d=2$ and the final transformation $T^{(n_l)}$ maps the output of the penultimate layer to $\mathbb{R}$.

Much of the interest in deep neural networks stems from the fact that these models are known to be universal function approximators. In other words, given a continuous function $f$ on a compact set of $d$-dimensional space $f : \mathbb{R}^d \to \mathbb{R}^k$, there exists a neural network $\Phi$ with (at least) one hidden layer and a non-linear activation function that approximates it to any desired degree of precision \citep{cybenko1989approximation}. It is thanks to this property and to the development of powerful non-convex optimization algorithms \citep{kingma2017adam} that neural networks have been extensively applied in a variety of fields where datasets in the form of input-output pairs are available.


Recent work has demonstrated that neural networks can be employed to represent complex signals including time series, images, and 3D scenes with a high level of detail \citep{nerf,sitzmann2019siren,tancik2020fourier}. In particular, standard MLPs can be trained to map a generic input coordinate $\vec{x}$ to the corresponding signal value at that coordinate. In our present application to strong lensing, this means pixel values of an observed image (see Fig.~\ref{fig:exampleSL}). Given that such a neural network is defined over a continuous domain, the ability of the resulting model to capture fine details of an image is not limited by any pre-defined grid resolution. Its accuracy depends only on the capacity and expressive power of the network architecture.

In the most general case, the target signal we aim to represent with a network is not directly available, but instead only implicitly defined. For instance, we might only have access to the gradients or higher-order derivatives of the image we wish to recover. As a result, following \citet{sitzmann2019siren}, we can frame the problem of implicit function learning as the search for $\Phi(\vec{x})$ subject to 
\begin{align}
    \label{eq:nn_constraint}
    \mathcal{C}_{m}\Big(\vec{a}(\vec{x}), \Phi(\vec{x}), \nabla \Phi(\vec{x}), \ldots\Big) = 0 \ ,
\end{align}
where $\{\mathcal{C}_{m}\}_{m=1}^M$ are a set of $M$ constraints that each relates the function $\Phi$ and/or its derivatives to a set of known quantities $\vec{a}(\vec{x})$. Each constraint, depending on its desired effect, applies to its own subset of the input domain $\Omega_m$ such that $\mathcal{C}_m=0$ holds for all $\vec{x} \in \Omega_m$. Implicit function learning can therefore be translated into the minimization of a loss function penalizing deviations of the collection of constraints from zero:
\begin{align}
    \label{eq:loss_general}
    L=\int_{\Omega} \sum_{m=1}^{M} \vec{1}_{\Omega_{m}}(\vec{x})\,\big\|\,\mathcal{C}_{m}\left(\vec{a}(\vec{x}), \Phi(\vec{x}), \nabla \Phi(\vec{x}), \ldots\right)\big\|\, \mathrm{d}\vec{x} \ ,
\end{align}
for a suitably chosen norm $\|\cdot\|$. The indicator function $\vec{1}_{\Omega_{m}}(\vec{x})$ restricts each constraint only to its domain, giving $1$ when $\vec{x} \in \Omega_m$ and $0$ otherwise.

\begin{figure*}
    \centering
    \includegraphics[width=\textwidth]{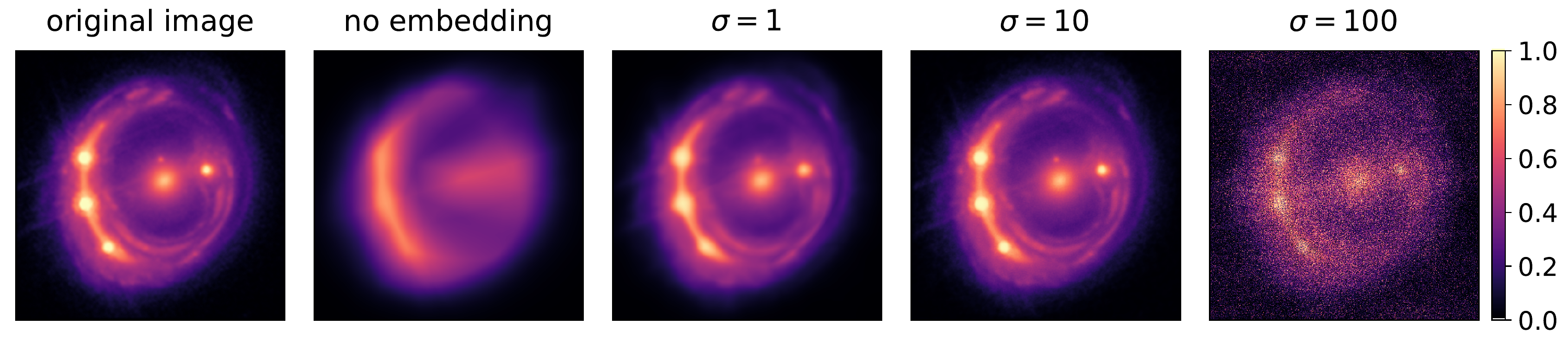}
    \caption{Illustration of the effect of scale $\sigma$ on fitting an image using a neural network including the Fourier feature mapping $\gamma_\sigma$. The original image (flux-normalized as in Fig.~\ref{fig:exampleSL}) is shown on the far left, and the remaining images show fully-connected network representations with varying $\sigma$ values for the embedding. Lower values (including no embedding) give smoother representations but fail to capture high spacial frequencies in the image, while high values tend to overfit the fine details and reconstruct the noise. The same network architecture and training scheme were used to produce all of the fits.}
    \label{fig:sigma}
\end{figure*}

In the simple case of fitting an image $\mathcal{I}(\vec{x})$ directly from its pixel values, a single constraint is present, namely $\mathcal{C}\Big(\mathcal{I}(\vec{x}), \Phi(\vec{x})\Big)=\mathcal{I}(\vec{x}) - \Phi(\vec{x})$, where $\vec{a}(\vec{x})=\mathcal{I}(\vec{x})$. The loss in this case assumes the form
\begin{align}
    \label{eq:loss_image_fit}
    L = \frac{1}{N} \sum_{i=1}^{N} \Big[ \mathcal{I}(\vec{x}_i) - \Phi(\vec{x}_i) \Big]^2,
\end{align}
where $N$ is the number of pixels, and we have chosen, for example, the mean squared error as the norm, as was used to produce Fig.~\ref{fig:exampleSL}. In more complex scenarios, the number and types of constraints depend on the details of the problem and on the amount of information available to train the model. In this work, these constraints arise directly from strong lensing physics, namely, the lens equation and its relation to the lens potential (see Sect.~\ref{sec:strong_lensing}).

A limitation of standard MLPs has been pointed out by \citet{rahaman2019spectral}, where it was shown that these models are biased towards low-frequency solutions. Where reconstructing the fine-grained details of an image is important, such as in representing realistic and complex natural images, this property can hinder the applicability of MLP networks. A number of solutions have been proposed to overcome this issue \citep{sitzmann2019siren, tancik2020fourier}. In this work, we implement the approach of \citet{tancik2020fourier}, in which the input coordinate vector is first passed through a Fourier feature mapping $\gamma_{\sigma}$ before being fed into the network:
\begin{align}
    \label{eq:embedding}
    \gamma_{\sigma}(\vec{x}) = \big[\cos (2 \pi \mat{A}\vec{x}),\, \sin (2 \pi \mat{A}\vec{x})\,\big]^\top \ .
\end{align}
Each entry in $\mathbf{A} \in \mathbb{R}^{n \times d}$ is sampled from $\mathcal{N}\left(0, \sigma^{2}\right)$, and the $\cos$ and $\sin$ functions act element-wise on their arguments. The input $\vec{x}$ is therefore mapped to $2n$ components via $\gamma_{\sigma}$, and $n$ is typically chosen to be larger than $d$. This straightforward operation has proven to effectively limit the spectral bias of standard neural networks and to enable more control over the output frequencies. Applying $\gamma_{\sigma}$ can thus be thought of as a pre-processing step whereby Eq.~(\ref{eq:mlp}) becomes $\Phi(\vec{x}) \rightarrow \Phi(\gamma_{\sigma}(\vec{x}))$.

\subsection{Hybrid differentiable lensing model with a neural network potential}

We model the lensing potential $\psi(\vec\theta)$ with a neural network $\Phi(\vec{x})$ as defined by Eq.~(\ref{eq:mlp}) and including the $\gamma_\sigma$ mapping. It is parameterized by a collection of weights and biases, which we denote by $\vec{\eta}_{\Phi}$, and $\vec{x}\in\mathbb{R}^2$ is a point in the (continuous) lens plane. The network has been implemented as a module within the \herculens code \citep{Galan2022} using Flax \citep{flax2020github}, a JAX-compatible library for neural network development. Since the model is fully differentiable, training the network (i.e. optimizing $\vec\eta_\Phi$) can be carried out using gradient descent optimization. It can also be done simultaneously with the optimization of other model components, such as analytic smooth lens parameters or source light parameters.

Supervised learning for deep neural networks requires a large number of labelled input-output pairs. In strong lensing applications, this typically means training on tens or hundreds of thousands of example images (outputs) generated from known input lensing parameters (inputs). In our case, the training set is instead comprised of the set of all pixel positions and their corresponding brightness values as provided in the original data image. No other prior training set needs to be created for our hybrid model to learn its implicit representation of $\psi$. We note, though, that the learned potential is specific to the data being modelled and does not necessarily generalize to other observations.

Given the observed image data $\vec{d}$ of a strongly lensed system, Eq.~(\ref{eq:nn_constraint}) can be written as the single constraint
\begin{align}
    \label{eq:constraint_SL}
    \mathcal{C}(\vec{d}, \nabla\Phi)=\vec{d}-\convop\lensingop_{\psi}(\nabla\Phi)\vec{s} \ ,
\end{align}
where, as described in Sect.~\ref{sec:strong_lensing}, we assume that the lens light $\vec{\ell}$ has been accurately modeled beforehand and subtracted from the data. Because the lens equation maps image to source plane coordinates via the gradient of the lens potential, we write the lensing operator $\lensingop_{\psi}$ here as an explicit function of $\nabla\Phi$.

Combining Eqs.~(\ref{eq:loss_general}) and (\ref{eq:constraint_SL}) while taking into account noise gives rise to the loss function
\begin{align}
    \label{eq:loss_SL}
    L(\vec{\eta})= \frac{1}{2}(\vec{d}-\vec{m})^{\top} \mathbf{C}_{\vec{d}}^{-1}(\vec{d}-\vec{m}) \ ,
\end{align}
where $\mathbf{C}_{\vec{d}}$ is the covariance matrix of the data, and we have defined the model
\begin{align}
    \label{eq:model}
    \vec{m}(\vec{\eta}) \equiv \convop\lensingop_{\psi}(\vec{\eta}_{\Phi},\vec{\eta}_{\psi})\vec{s}(\vec\eta_{s}) \ .
\end{align}
Equation (\ref{eq:model}) is the general form of our model, where $\vec\eta_s$ are the source light parameters (whether the source is analytic or pixelated), and $\vec\eta$ refers to the collection of all model parameters. We distinguish $\vec\eta_{\psi}$ from $\vec\eta_{\Phi}$, where the former refer to parameters of an analytic smooth potential, which can be included when $\Phi$ is only modelling perturbations. Without loss of generality, we assume the noise to be uncorrelated, which makes $\mathbf{C}_{\vec{d}}$ diagonal. Equation (\ref{eq:loss_SL}) can be understood as the negative log-likelihood of normally distributed data, and is similar to Eq.~(\ref{eq:loss_image_fit}) but with the mean squared error replaced by the $\chi^2$.

A schematic illustration of our modeling approach is shown in Fig.~\ref{fig:model}. The resulting framework consists of a \emph{hybrid model}, where the neural network lens potential operates seamlessly within a differentiable strong lensing simulator. Our main goal is to leverage a general-purpose neural network to accurately model the potential without any specific inductive biases on its form and with the only constraints stemming from the physics of strong lensing and the observed data.

The use of implicit neural networks to model $\psi$ also presents some immediate advantages with respect to pixelated models of the lensing potential: once trained, the network can be queried at any continuous input point without the need to interpolate between pre-defined grid positions. We note as well that owing to the differentiable nature of $\Phi$, the spatial gradients of the modelled potential can be efficiently computed via auto-differentiation with respect to the input $\vec{x}$. We therefore have direct access to the map of deflection angles $\vec\alpha(\vec\theta)$, which can be used as an additional test of the method's accuracy.

\begin{figure*}
    \centering
    \includegraphics[width=0.24\textwidth]{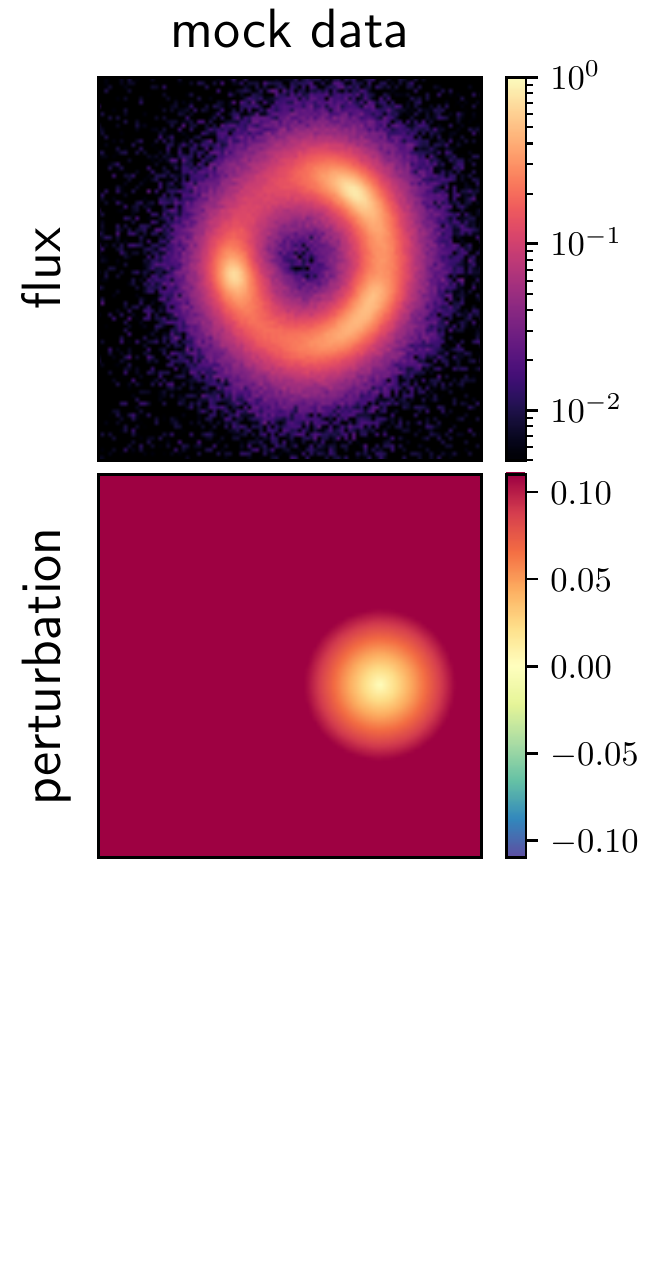}
    \includegraphics[width=0.735\textwidth]{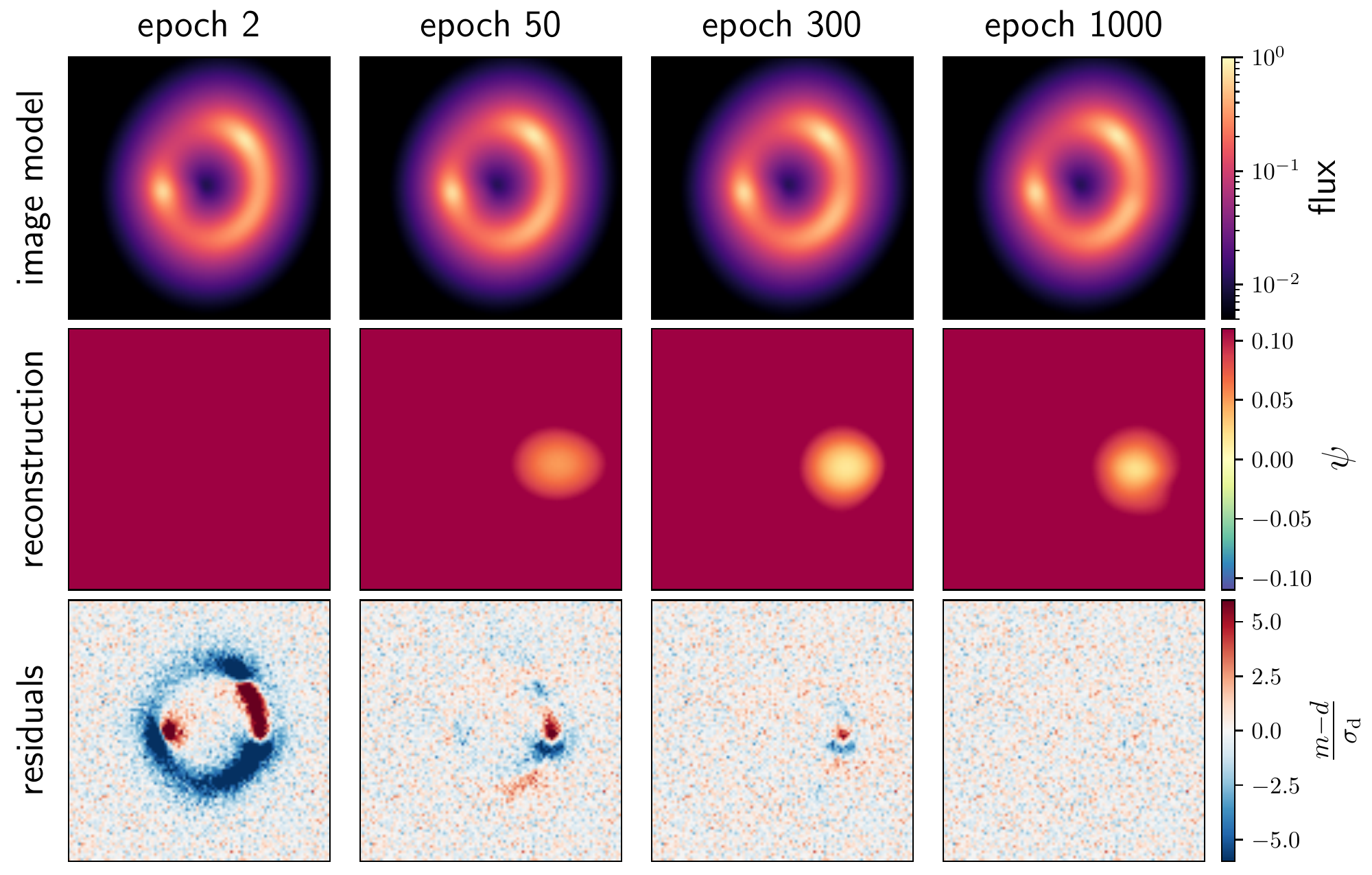}
    \caption{Evolution of the network output prediction when modeling an example localized perturbation to a smooth lens potential. Mock noisy lensing image data and the true perturbation are shown on the left. The predicted lensing image, the network reconstruction of $\psi_\mathrm{pert}$, and residuals between the data and the predicted image at different training epochs are also shown. Training is performed via back-propagation of the gradients through the simulator, as presented in Fig.~\ref{fig:model}.}
    \label{fig:training_progress}
\end{figure*}

\begin{figure}
    \centering
    \includegraphics[width=0.48\textwidth]{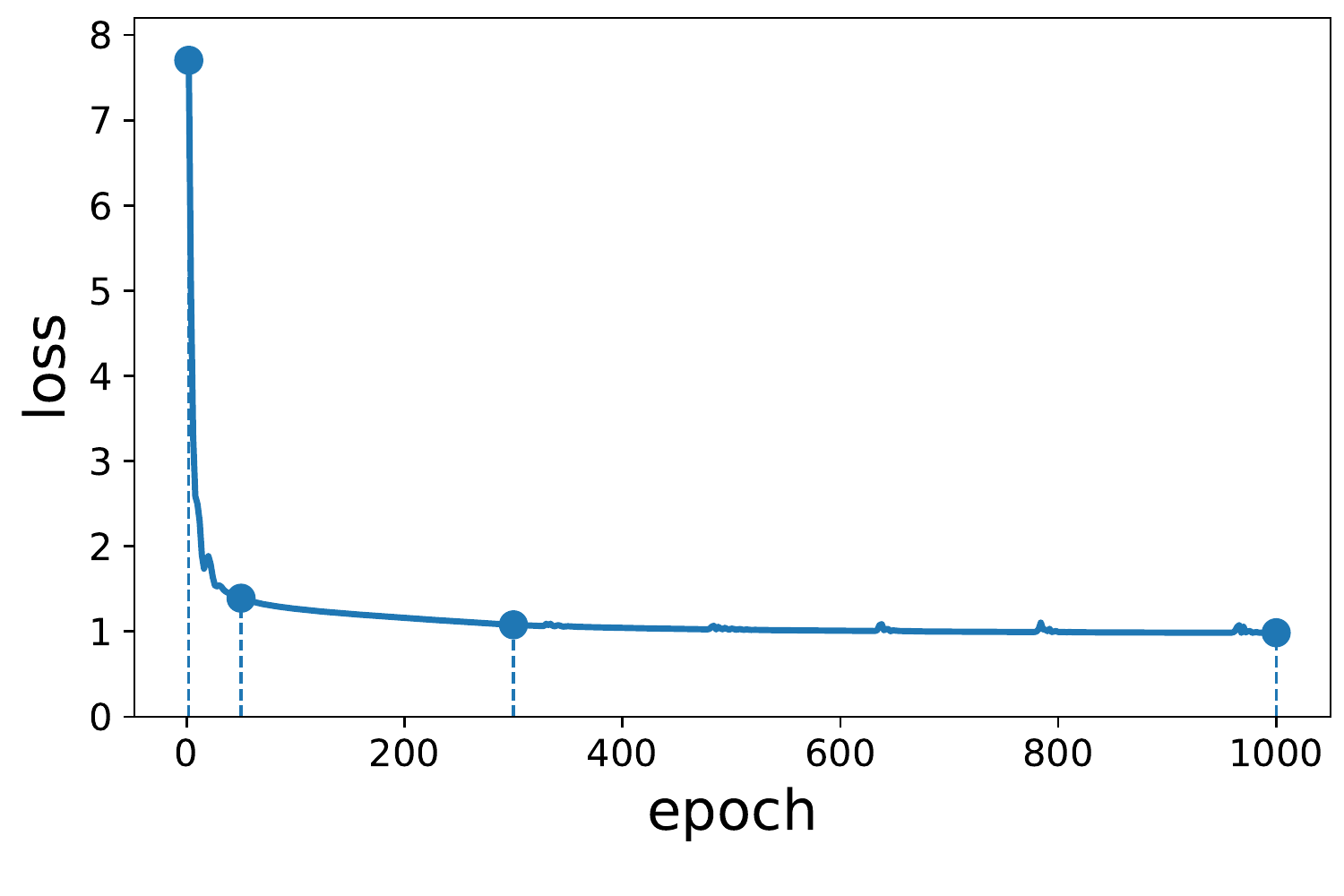}
    \caption{Evolution of the loss function during the training of $\psi_{\rm pert}$ in the example of Fig.~\ref{fig:training_progress}. Epochs for which the image model, the reconstructed perturbation, and image plane residuals are shown in Fig.~\ref{fig:training_progress} are indicated on the plot. By epoch 300, the network already produces a good fit to the data with a small loss value and residuals largely consistent with noise.}
    \label{fig:training_loss}
\end{figure}

\subsection{Network architecture and parameter optimization}

Throughout our study, we keep the complexity of the architecture relatively low, using 5 hidden layers of 100 nodes each to specify $\Phi(\vec{x})$. The activation $\phi$ was chosen to be the Swish function \citep{ramachandran2017searching}, which we found to outperform the common rectified linear unit (ReLU) in terms of the quality and smoothness of the reconstructed potential gradients (i.e. deflection angles).

We chose $m=100$ for the dimension of the Fourier embedding [Eq.~(\ref{eq:embedding})] and a standard deviation $\sigma=0.05$. We recall that this mapping is not a trained parameter and serves only to focus the network towards a certain range of spatial frequencies. The value we chose was to prevent the network from fitting the noise and provided good results across all of the experiments carried out. Apart from the choice of activation function, we did not fine-tune all hyperparameters, a procedure that would most likely improve results, but which is beyond our present scope.

We illustrate the effect of varying $\sigma$ on fitting an image of a strongly lensed quasar system in Fig.~\ref{fig:sigma}. The same data and network architecture as were used in Fig.~\ref{fig:exampleSL} were used here, namely a simple MLP fitting the observed flux of RX\,J1131$-$1231. Low values of $\sigma$ (as well as no embedding at all) lead the network to only capture the low spatial frequencies of the image, meaning the resulting fit is smoother than the original while noise is effectively filtered out. This is in agreement with the recent finding of \citet{rahaman2019spectral} that neural networks are intrinsically biased to learn low frequency functions. On the other hand, higher $\sigma$ values can result in overfitting the data, so that $\Phi$ learns to reproduce the noise in addition to the underlying signal. This example shows how $\sigma$ acts as a tunable knob for the network allowing control over the desired output level of detail.

In Sect.~\ref{sec:results}, we perform experiments where we model different combinations of lens potential perturbations and the full potential directly. In all the experiments, we train our network using the Adam optimizer \citep{kingma2017adam} with learning rate $10^{-3}$ as implemented in the Optax framework \citep{optax2020github}, a gradient processing and optimization library for JAX.

As an illustration of the typical training procedure of $\Phi$, we consider the concrete example of modeling a single localized sub-halo perturbation to a smooth potential. Progress of $\Phi$ towards the true $\psi_\mathrm{pert}$ as a function of epoch is shown in Fig.~\ref{fig:training_progress}, with the only training signal coming from the loss in Eq.~(\ref{eq:loss_SL}). At the start of training, the network is randomly initialized, and thus its output is far from the ground truth (e.g. at epoch 2). As training progresses, the model finds the right location and scale of the perturbation, obtaining a good fit after only around 300 epochs, as shown in Fig. \ref{fig:training_loss}.

\begin{figure*}
    \centering
    \includegraphics[width=\textwidth]{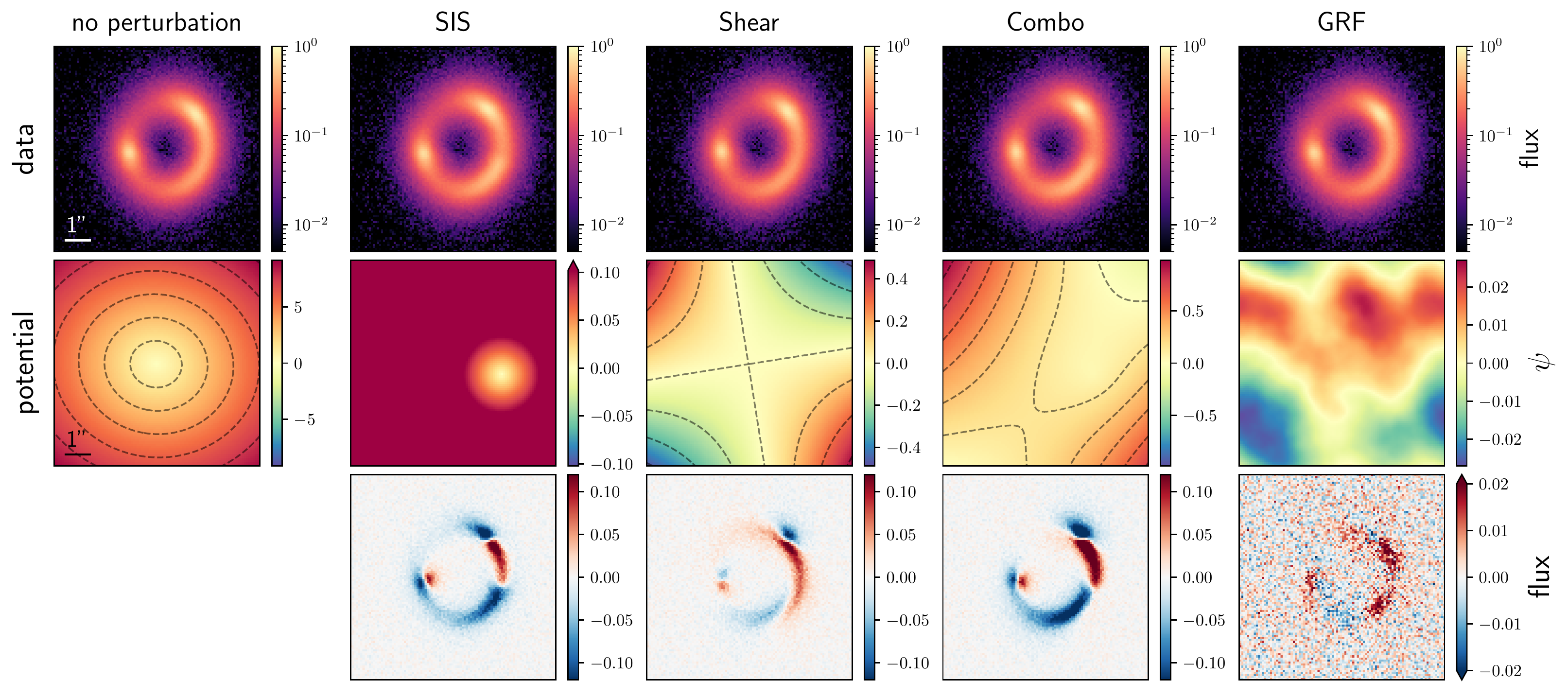}
    \caption{Mock data modelled in this work. The leftmost column shows the image and lensing potential of a mock system without any perturbations, while subsequent columns show the mock image (first row) of the same system under the influence of a different perturbing potential (second row). The difference with the unperturbed case is shown in the bottom row. Details on generating the mocks and the parameters of the system and perturbations are given in Sect. \ref{sec:experiments} and in Tables \ref{tab:perturbation_fits} and \ref{tab:sie_fits}. The Combo case is simply a combination of the SIS and Shear perturbations. Contours are shown for the Shear and Combo cases to better trace the shape of the lensing potential. We also note that the GRF perturbing potential has 2-5 times lower amplitude than the other ones.}
    \label{fig:mock_data}
\end{figure*}

\section{Experimental setup}
\label{sec:experiments}
Any lens modeling effort consists of three main components: the brightness distributions of the source and lens, and the mass distribution of the lens that is linked to the lens potential. 
None of these components is precisely known, and analyzing any lensing system relies on priors, the extent to which depends on the method used.
In addition, all components can have detailed structure on small scales that cannot be fully captured by the typically used macroscopic profiles, like SIE and S{\'e}rsic \citep{Sersic1963}.
For example, there is a known but difficult to capture degeneracy between source brightness and lens potential that can complicate any attempt to model the system down to the noise \citep[e.g. see][]{Vernardos2022}.

In this work, we explore how model-agnostic neural networks can provide a reconstruction of the lens potential across all scales with only little prior information.
Therefore, in the mock data used in our experiments we assume a smooth, analytic, elliptical S{\'e}rsic brightness profile for the source light, a SIE profile as the main lens potential, and no light from the lens (the choices of parameters for the analytic profiles used can be found in Tables \ref{tab:perturbation_fits} and \ref{tab:sie_fits} and in Sect.~\ref{sec:mock_data}).
We then perturb the main lens potential in different, physically motivated ways, which we explain in detail below.
We note that the experimental setup is very similar to \citet{Galan2022}, and we therefore focus on emphasizing the differences here.

\subsection{Single localized subhalo}
We simulate a single, localized, spherical DM subhalo, assumed to have an isothermal profile with Einstein radius $\theta_{\rm E,sub} = 0\farcs07$ (for reference, the main deflector has $\theta_{\rm E} = 1\farcs8$).
Assuming typical redshifts of Early-type/Early-type lensing systems \citep{OldhamAuger2018} for the lens and the source, $0.3$ and $0.7$ respectively, and that the perturbing subhalo is in the plane of the lens, the mass within $\theta_{\rm E}$ is $10^9$ M$_{\odot}$.
This is comparable to massive subhalos detected via their lensing effect that have been previously reported in the literature \citep[e.g.][]{Vegetti2010}.
The mean perturbation level computed within the region containing the lensed arcs corresponds to $6.1\%$ of the main lens potential.
The resulting perturbing lens potential is shown in Fig. \ref{fig:mock_data}.

\begin{figure*}
    \centering
    \includegraphics[width=\textwidth]{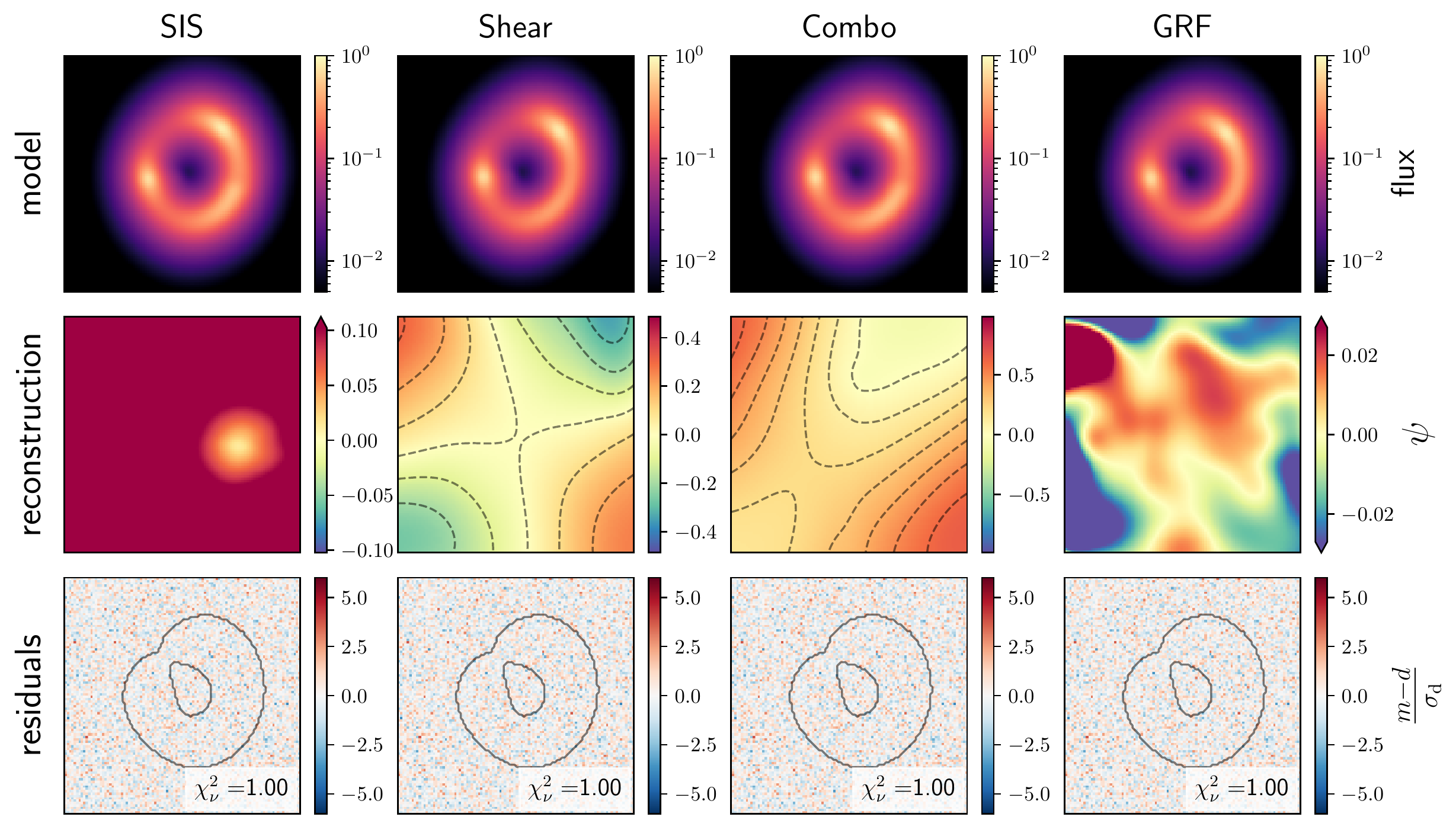}
    \caption{Top row: models of the mock data shown in Fig.~\ref{fig:mock_data}. Middle row: reconstructed perturbations. Bottom: model residuals in normalised units, i.e. divided by the diagonal, $\sigma_{\rm d}$, of the data covariance matrix from Eq. (\ref{eq:loss_SL}). We note in particular that within the area of the most significant lensed flux, as indicated by the mask (solid contour), the residuals are identical to noise. The reduced chi-squared $\chi^2_{\nu}$ of the models is also given in the panels of the bottom row.}
    \label{fig:perturbations}
\end{figure*}

\subsection{Population of subhalos}
The net effect on the lens potential of a population of DM subhalos along the line-of-sight can be approximated by a 
Gaussian Random Field \citep[GRF,][]{ChatterjeeKoopmans2018,Bayer2018,Vernardos2020,Vernardos2022}.
The statistical properties of such a random field can be defined by its (isotropic) power-spectrum, which follows a power-law:
\begin{equation}
    \label{eq:def_grf}
    {\rm PS}(k) = A_{\rm GRF} \, k^{-\beta_{\rm GRF}} \, ,
\end{equation}
where $k$ is the wave-number, $\beta_{\rm GRF}$ is the power-law slope, and $A_{\rm GRF}$ is a normalizing factor that depends on $\beta_{\rm GRF}$ and the size of the field of view and is related to the variance of the GRF, $\sigma^2_{\rm GRF}$ \citep[for the exact formula, see][]{Chatterjee2019_thesis}.
The value of $\beta_{\rm GRF}$ determines  the distribution of power at each length scale: a large value leads to extended and smooth variations, whereas a small value creates a large number of localized and grainy structures.

Typical ranges for $\sigma_{\rm GRF}$ and $\beta_{\rm GRF}$ that have been explored in the literature are $\sigma_{\rm GRF} \in [10^{-5},10^{-2}]$ and $\beta_{\rm GRF} \in [3,8]$ \citep{Chatterjee2019_thesis,Vernardos2020}, while \citet{Bayer2018} exclude a GRF variance larger than $\sigma^2_{\rm GRF}=10^{-2.5}$, based on HST observations of the strong lens system SDSS\,J0252$+$0039.
In this work, we set $\mathrm{log}_{10}A_{\rm GRF}=-8$ (i.e. $\sigma_{\rm GRF}^2 \approx 10^{-3.8}$) and $\beta_{\rm GRF}=-3$, such that it leads to a GRF that is not unphysically large and contains both small and large-scale features. The resulting GRF realization is shown in Fig. \ref{fig:mock_data}.
The mean perturbation level computed within the region containing the lensed arcs is about $0.4\%$ per cent of the main lens potential.

\subsection{External shear}
Perturbations on the same scale as the typically used smooth profiles (that is, scales comparable to the Einstein radius $\theta_{\rm E}$) can occur in a lensing galaxy due to higher order moments in its mass distribution.
For example, there can be ellipticity gradients \citep{VandeVyvere2022_TDC7}, the presence of a bar \citep{Hsueh2018}, or deviations from ellipticity that can be attributed to multipolar azimuthal perturbations \citep{VandeVyvere2022_TDC7}.
In the latter case, \citet{Galan2022} consider the effects of an octupole; instead, here we consider the more common case of a quadrupole through an external shear perturbation, which simulates first-order effects of massive structures in the vicinity of the lens.
This moment is widely used to describe lensing deflections at distances much larger than the Einstein radius, usually induced at the location of a given lens by mass concentrations (e.g. galaxies) that lie close to it and perpendicular to the line of sight.
Although this external shear lens potential component has an analytical expression, we choose to treat it agnostically as a perturbation. 
Motivated by typical values found in the literature, we choose a shear magnitude $\gamma_{\rm ext}=0.032$ and direction $\phi_{\rm ext}=144^{\circ}$. The corresponding perturbing lens potential is shown in Fig.\ref{fig:mock_data} and it amounts to 1.1 \% of the main lens potential within the region containing the lensed arcs.

\subsection{Generating the mocks}
\label{sec:mock_data}
We simulate data with the independent code \molet \citep{Vernardos2021} in order to prevent the occurrence of potentially advantageous minima during optimization. This choice also better mimics real-world situations, as the data never exactly corresponds to any model generated by the modeling code itself.

We consider typical observations of strongly lensed galaxies as observed with HST and the Wide Field Camera 3 (WFC3) instrument, in the near infrared (F160W filter). The pixel size is $0.08$ arcsec, and the field of view is $8 \times 8$ arcsec${}^2$. We do not explore effects due to incorrect PSF modeling, hence for simplicity we use a Gaussian PSF with $0.3$ arcsec full-width-at-half-maximum. We consider an exposure time of 9600 seconds, which corresponds approximately to 4 HST orbits. The magnitude zero-point is 25.9463 mag, the sky brightness is 22 mag/arcsec$^2$, and the detector readout noise is 4 electrons, in line with the assumed instrument and filter \citep{Gennaro2018wfc3}. The simulated noise contains both a Gaussian component from readout noise and Poisson component from shot noise.

The source light is taken as a S{\'e}rsic profile, with half-light radius $\theta_{\rm eff}=0.8$, S{\'e}rsic index $n_{\rm Sersic}=2$, axis ratio $q=0.82$, and orientation angle $\phi=170^{\circ}$. The center of the distribution is located in the source plane at $(0.40, 0.15)$ arcsec, resulting in the lensed annulus seen in the image plane. We use the same parameter definitions as in \citet{Galan2022} and therefore refer the reader to their Appendix B.

These assumptions for generating the mock data result in a sufficiently realistic experimental setup in which we can evaluate our method.
The first column of Fig.~\ref{fig:mock_data} shows the image of an unperturbed simulated system, while the remaining columns show mock data corresponding to the different perturbation cases to which we apply our method.

\begin{figure*}
    \centering
    \includegraphics[width=\textwidth]{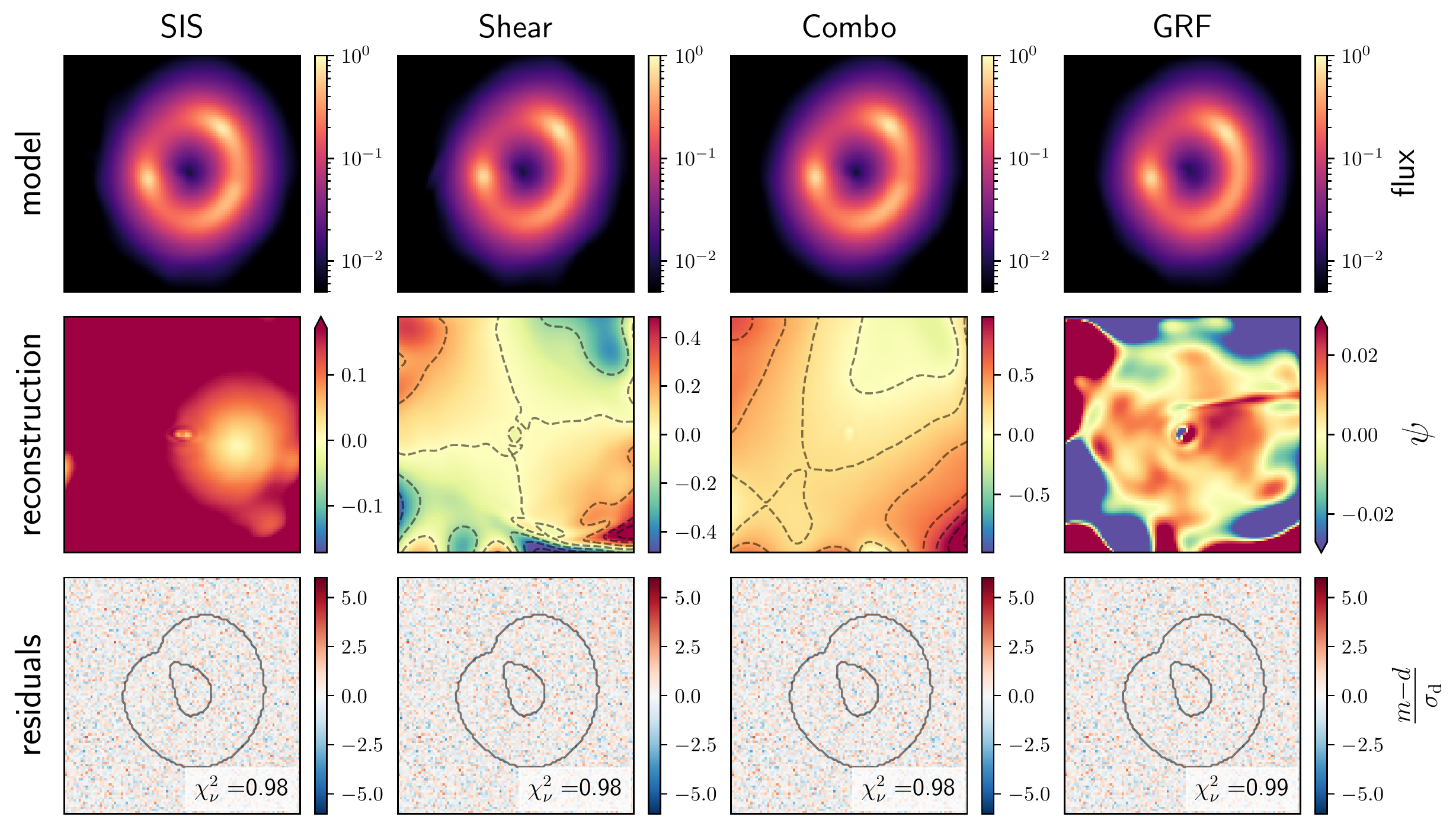}
    \caption{Same as Fig.~\ref{fig:perturbations}, but in this case the full potential was modelled and then the true main SIE was subtracted in order to obtain the reconstructions shown here. We changed the amplitude scale of the reconstructed SIS (left column) to better outline some detailed structure (artifacts) that appear for this model.}
    \label{fig:full}
\end{figure*}

\section{Results}
\label{sec:results}
We analyze the mock data described above with our neural network-based method to model the lens potential.
Our objective is to recover the perturbations to the main lens potential as accurately as possible without any prior assumption on their form.
We also assess how well we can reconstruct the full lensing potential without assuming any analytical profile.
In all our experiments, we assume perfect knowledge of the source.

\subsection{Recovering perturbations to the smooth potential}
In this case, the main lens potential is fixed and our continuous neural field model is used to discover a hidden perturbation for which little or no prior information is available.
This situation can commonly occur when a smooth model is able to fit the largest-scale features of a system, but as it does not explicitly account for perturbations, localized features are present in the residuals.

Figure \ref{fig:perturbations} shows the model, reconstructed perturbations, and residuals for the four different mock observations that we examine.
Visually, the residuals are very close to the noise level (the reduced chi-squared $\chi^2_{\nu}$ is unity), and the reconstructed perturbation fields very closely match the truth.
In the case of the localized SIS substructure, the model is able to recover the correct shape, magnitude, and location of the perturbation.
Similarly, the shape and amplitude of the extended Shear and Combo perturbations are very close to the truth.  
For the GRF perturbation, the network is able to retrieve its general properties and detailed features, but small artifacts appear near the edges of the field of view.
In fact, we should expect the fidelity of the reconstructions to decrease the further we are from the gravitationally lensed flux of the source, roughly indicated by the mask in the second row of Fig. \ref{fig:perturbations}, due to the lack of signal.
Despite the apparent lack of signal, our method also reconstructs the perturbations outside the mask remarkably well, even in the case of the complex GRF, albeit with less accuracy.

To further evaluate the quality of the reconstructions, we fit analytic SIS and external shear profiles to the reconstructions of the SIS, Shear, and Combo models.
The results are listed in Table \ref{tab:perturbation_fits}, where we see that the true value of the parameters is always within the $1 \sigma$ confidence interval of our fits.
In a similar fashion, we calculate the amplitude and slope of the power spectrum of the GRF reconstruction.
This is done within the mask (as well for the true underlying GRF), where we find systematic errors of only a few per cent.

\subsection{Modeling the full potential}
We also model the same mock data without any knowledge of the main, smooth lens potential component of the lens, namely without assuming an SIE profile.
We expect this less constrained version of the problem to be more degenerate and harder to optimize, as our neural network method must now account for both an extended smooth potential and local perturbations.
However, based on the Combo case examined above, which does combine extended and local perturbations albeit in a limited way, we don't expect this entirely to fail.

Figure \ref{fig:full} shows the model, reconstructed perturbations after subtracting the true SIE main potential, and residuals for the four different data cases.
As when fitting only the perturbations, the residuals here are visually very close to the noise level, with an indication of slight overfitting from the $\chi^2_{\nu}$ being below unity.
The reconstructed fields for the SIS, Shear, and Combo cases closely match the true perturbations, although artifacts outside the mask are amplified, in particular around the edges and in the center of the field-of-view.
This is also supported by the analytic fits, for which parameters are listed in the last column of Table \ref{tab:perturbation_fits}, and which are almost always within $1 \sigma$ confidence interval from the truth.
The reconstructed GRF perturbation visually bears little resemblance to the true field, although its power spectrum is recovered remarkably well, and its fitted power-law parameters have a systematic bias of just a few percent.

Having fitted the full potential, we can also investigate how well the main smooth potential is reconstructed.
We do so by fitting an analytic SIE profile directly to the full reconstruction (meaning the perturbations are included).
The fitted parameters are listed in Table \ref{tab:sie_fits} and show small systematic biases of a few percent, except for the ellipticity and position angle in the Shear and Combo models.
This reflects the well-known degeneracy between these two parameters and the magnitude and orientation of the shear \citep[e.g., see Fig. 3 of][]{Vernardos2022}.
The presence of a massive subhalo in the SIS and Combo models also biases the value of $\theta_{\rm E}$ towards higher values, because more mass is directly added to the total lens potential.
This is not the case with the GRF perturbations, which should average to zero within the mask and whose amplitude is too low to bias $\theta_{\rm E}$ anyway.

\section{Discussion and conclusion}
\label{sec:conclusion}
In this work, we present a new way of using neural networks to address the problem of mass reconstruction in galaxy-galaxy gravitational lensing.
By being embedded in a model that explicitly satisfies the lens equation, our neural network is agnostic to any specific mass distribution and converges to the solution using automatically calculated derivatives.
The method is able to capture both large-scale (i.e., spanning the field-of-view) and small-scale (i.e., of the order of a few pixels) features in lensing potentials, which corresponds to roughly two orders of magnitude in terms of angular scale.

We are able to successfully model large-scale potentials, such as an external shear component or the combination of a primary smooth potential (e.g. SIE) and a perturbing field representing a single massive subhalo, an external shear, or a population of lower mass subhalos.
In the latter case, the parameters of the smooth potential are recovered in a subsequent fit with biases of only a few percent (including the effect of the perturbations), without using any analytical profile or other prior knowledge during the reconstruction (see Table \ref{tab:sie_fits}).

The parameters of the perturbing potentials are well recovered in the case of perfect knowledge of the main potential (see Fig. \ref{fig:perturbations} and Table \ref{tab:perturbation_fits}).
This is also true of the reconstruction in areas outside of the mask and away from where most of the lensed flux is observed. Given the much lower signal-to-noise ratio in these pixels, such accuracy is not necessarily anticipated. When reconstructing the full potential, it is primarily in these regions that artifacts appear.
Nevertheless, the reconstructions are still good enough without model residuals such that we can recover parameters of the perturbations with very small biases.

Regarding the case of a subhalo population, which is the most complicated but also the weakest perturbation we examined, its detailed structure is remarkably well reconstructed if perfect knowledge of the main potential is assumed (see the bottom right panel in Fig. \ref{fig:training_progress}).
On the other hand, assuming no knowledge of the underlying potential, the statistical properties of the population as described by its power spectrum are still very well recovered (see Table \ref{tab:perturbation_fits}), despite the actual reconstructed potential being qualitatively different from the truth.
A similar result was obtained by \citet{Vernardos2022}, who used the semi-linear inversion technique to reconstruct the perturbations.

Our method relies on the standard deviation $\sigma$ of the Fourier feature mapping (see Sect. \ref{sec:method} and Eq.~\ref{eq:embedding}) as a hyper-parameter.
As illustrated in Fig.~\ref{fig:sigma}, the higher its value, the finer the features (including noise) in the data that can be captured by the neural network.
This is in direct analogy to the regularization strength in pixelated models, like the semi-linear inversion technique \citep[e.g.][]{Vernardos2022} and when using wavelets \citep{Galan2021,Galan2022}.
In this work, we chose a value obtained by trial and error, which works well for analyzing the mock data and the given signal-to-noise explored here.
However, an interesting avenue for future research is to integrate this hyper-parameter into our differentiable inference scheme, and obtain its most probable value in a Bayesian way. We leave the development of such a framework for future work.

In this initial presentation of our new method, we examine lens configurations that have a smooth, analytical source brightness profile.
This is a deliberate choice due to the strong degeneracy known to exist between small-scale structure in the source light and the lens mass \citep[see][for a study of this degeneracy]{Vernardos2022}.
Systems with complicated source structure are expected to affect the signal-to-noise ratio per pixel and consequently the choice of $\sigma$.
In addition, we keep the source fixed to the truth and do not attempt to reconstruct it simultaneously with the lens potential.
Although this is clearly not a realistic scenario, it does allow us to validate our method in a more restricted parameter space and controlled set of experiments. We intend to explore both effects of a structured and unknown source in future work.

From the point of view of machine learning, improvements to the current model can be approached from several angles. First, the network design can be modified to allow for the quantification of uncertainty. In this way, its output would also include an estimate of the error associated with its prediction. This could be done, for example, by using the MC-Dropout technique as in \citet{Perreault_Levasseur_2017} to model the so-called epistemic uncertainty. One can also add an additional output node to the network in order to model the covariance of predictions and capture the so-called aleatoric uncertainty, as in \citet{fluri2019cosmological}. Another interesting direction would be to study ways to use better initialization schemes of the network weights, or use pre-training routines, followed by a fine-tuning phase for each new lensing potential. Finally, to the best of the authors' knowledge, this is one of the earliest applications of implicit representation learning to cosmology. In light of the extraordinary success achieved by modern methods designed within this framework, particularly for the synthesis of 3D images \citep{nerf}, we believe that the exploration of their application to cosmology represents an exciting avenue for future research.

In conclusion, we have presented a promising machine learning method applied for the first time to lens potential reconstruction problems.
Our method can accurately reconstruct, in a continuous way, single massive subhalos, complex subhalo populations, and large-scale potentials that span the entire field of view. For the latter, the simple external shear that we have explored here could be straightforwardly extended to include ellipticity gradients, twists, and higher-order moments \citep[e.g.][]{VandeVyvere2022_TDC7}.
Our experiments covered two extreme cases, both with successful results: either perfect knowledge or no knowledge of the main lens potential. This motivates the development of a more traditional approach that reconstructs the lens potential as a combination of an analytical profile for the main potential (e.g. SIE or power-law) and a free-form field that captures any remaining perturbations.
Given that our method is implemented within \herculens, we will explore this in future work, as well as degeneracies with the source surface brightness.

\renewcommand{\arraystretch}{1.4}
\begin{table*}
	\centering
	\caption{Evaluation of the reconstructed perturbing potentials (Sect. \ref{sec:results}) by fitting analytical profiles to the SIS, Shear, and Combo models, and calculating the power spectrum of the GRF model. For the Combo model, the first four parameters refer to the shear and the others to the SIS. The power spectrum parameters of the GRF were fitted within the mask, as were the true perturbations.}
	\label{tab:perturbation_fits}
	\begin{tabular}{rrrrrrrr}
		&name & units & truth & $\psi_\mathrm{pert}$ only & $\psi_\mathrm{sm} + \psi_\mathrm{pert}$ \\

\hline
\multirow{3}{*}{\begin{sideways}SIS\end{sideways}}
 & $x_{0,\rm sub}$ & arcsec & $1.9$ & $1.940_{-0.249}^{+0.270}$ & $1.950_{-0.252}^{+0.284}$ \\
 & $y_{0,\rm sub}$ & arcsec & $-0.4$ & $-0.411_{-0.221}^{+0.218}$ & $-0.395_{-0.229}^{+0.194}$ \\
 & $\theta_{\rm E,sub}$ & arcsec & $0.07$ & $0.069_{-0.004}^{+0.004}$ & $0.069_{-0.004}^{+0.003}$ \\
\hline
\multirow{4}{*}{\begin{sideways}Shear\end{sideways}}
 & $x_{0,\rm ext}$ & arcsec & $0.0$ & $0.006_{-0.138}^{+0.134}$ & $-0.001_{-0.141}^{+0.130}$ \\
 & $y_{0,\rm ext}$ & arcsec & $0.0$ & $0.013_{-0.135}^{+0.136}$ & $0.004_{-0.145}^{+0.128}$ \\
 & $\gamma_{\rm ext}$ & $-$ & $0.032$ & $0.032_{-0.002}^{+0.002}$ & $0.032_{-0.002}^{+0.002}$ \\
 & $\phi_{\rm ext}$ & $^\circ$ & $144.0$ & $144.126_{-2.793}^{+2.258}$ & $143.913_{-2.779}^{+2.494}$ \\
\hline
\multirow{7}{*}{\begin{sideways}Combo\end{sideways}}
 & $x_{0,\rm ext}$ & arcsec & $0.0$ & $0.085_{-0.458}^{+0.507}$ & $0.078_{-0.391}^{+0.443}$ \\
 & $y_{0,\rm ext}$ & arcsec & $0.0$ & $-0.060_{-0.351}^{+0.304}$ & $-0.033_{-0.316}^{+0.270}$ \\
 & $\gamma_{\rm ext}$ & $-$ & $0.032$ & $0.031_{-0.002}^{+0.002}$ & $0.031_{-0.002}^{+0.002}$ \\
 & $\phi_{\rm ext}$ & $^\circ$ & $144.0$ & $143.536_{-3.045}^{+2.960}$ & $143.637_{-2.798}^{+2.905}$ \\
 & $x_{0,\rm sub}$ & arcsec & $1.9$ & $1.749_{-0.483}^{+0.462}$ & $1.812_{-0.481}^{+0.523}$ \\
 & $y_{0,\rm sub}$ & arcsec & $-0.4$ & $-0.248_{-0.770}^{+0.816}$ & $-0.275_{-0.699}^{+0.735}$ \\
 & $\theta_{\rm E, sub}$ & arcsec & $0.07$ & $0.070_{-0.004}^{+0.004}$ & $0.070_{-0.004}^{+0.004}$ \\

        \hline
        \multirow{2}{*}{\begin{sideways}GRF\end{sideways}}
        & $\log_{10}A_{\rm GRF}$  &  $-$  & $-8.231 ^{+ 0.005 }_{- 0.005 }$ & $-8.242 ^{+ 0.005 }_{- 0.005 }$ & $-8.088 ^{+ 0.001 }_{- 0.001 }$ \\
        & $\beta_{\rm GRF}$          &  $-$  & $-3.044 ^{+ 0.019 }_{- 0.019 }$ & $-3.106 ^{+ 0.02 }_{- 0.02 }$   & $-3.135 ^{+ 0.004 }_{- 0.004 }$ \\
	\end{tabular}
\end{table*}

\renewcommand{\arraystretch}{1.4}
\begin{table*}
	\centering
	\caption{Smooth potential parameters obtained by fitting an SIE profile to the $\psi_\mathrm{sm} + \psi_\mathrm{pert}$ models. We note that the full reconstructed potential (including the perturbations) is fitted.}
	\label{tab:sie_fits}
	\begin{tabular}{rrrrrrrr}
		& name & units & truth              & SIS & Shear & Combo & GRF \\
        \hline
 & $x_0$ & arcsec & 0.0 & $0.124_{-0.009}^{+0.009}$ & $0.036_{-0.009}^{+0.010}$ & $0.126_{-0.010}^{+0.009}$ & $0.043_{-0.009}^{+0.010}$ \\
 & $y_0$ & arcsec & 0.0 & $0.031_{-0.008}^{+0.008}$ & $0.038_{-0.008}^{+0.008}$ & $0.026_{-0.008}^{+0.007}$ & $0.045_{-0.008}^{+0.008}$ \\
 & $q_{\rm m}$ & $-$ & 0.73 & $0.729_{-0.014}^{+0.014}$ & $0.626_{-0.011}^{+0.012}$ & $0.627_{-0.010}^{+0.011}$ & $0.737_{-0.015}^{+0.014}$ \\
 & $\phi_{\rm m}$ & $^\circ$ & 82.0 & $83.583_{-1.252}^{+1.263}$ & $72.790_{-1.136}^{+1.043}$ & $71.460_{-1.055}^{+1.119}$ & $84.253_{-1.441}^{+1.424}$ \\
 & $\theta_{\rm E}$ & arcsec & 1.6 & $1.678_{-0.003}^{+0.003}$ & $1.612_{-0.003}^{+0.004}$ & $1.691_{-0.003}^{+0.003}$ & $1.600_{-0.003}^{+0.003}$ \\

	\end{tabular}
\end{table*}

\begin{acknowledgements}
GV has received funding from the European Union’s Horizon 2020 research and innovation programme under the Marie Sklodovska-Curie grant agreement No 897124.
This research was made possible by the generosity of Eric and Wendy Schmidt by recommendation of the Schmidt Futures program.
This programme is supported by the Swiss National Science Foundation (SNSF) and by the European Research Council (ERC) under the European Union’s Horizon 2020 research and innovation programme (COSMICLENS: grant agreement No 787886).
\end{acknowledgements}


\bibliographystyle{aa}
\bibliography{biblio}


\end{document}